\documentclass{book}
\usepackage{makeidx,epsfig}
\usepackage{setspace,graphicx}
\usepackage{Generic}
\makeindex
\begin{document}
%
%


\title{ The dark secrets of gaseous nebulae\\ -- highlights from deep spectroscopy }
\author{ Xiaowei Liu \\ Professor of Astronomy \\ Kavli Institute for Astronomy and Astrophysics, Peking University  }

\begin{doublespace}
\maketitle
\pagenumbering{arabic}

\chapter{The dark secrets of gaseous nebulae}

\section{Emission line nebulae}

The existence and distribution of the chemical elements and their isotopes is a
consequence of nuclear processes that have taken place in the past in the Big
Bang and subsequently in stars and in the interstellar medium (ISM) where they
are still ongoing (Pagel, 1997).  A large body of our knowledge of the
distribution and production of elements in the universe rests on observations
and analyses of photoionized gaseous nebulae.  Ionized and heated by strong
ultraviolet (UV) radiation fields, photoionized gaseous nebulae glow by
emitting strong emission lines (Osterbrock and Ferland, 2005). They are
therefore also commonly named emission line nebulae.  

\begin{figure}
\centering

\centerline{\includegraphics[width=120mm]{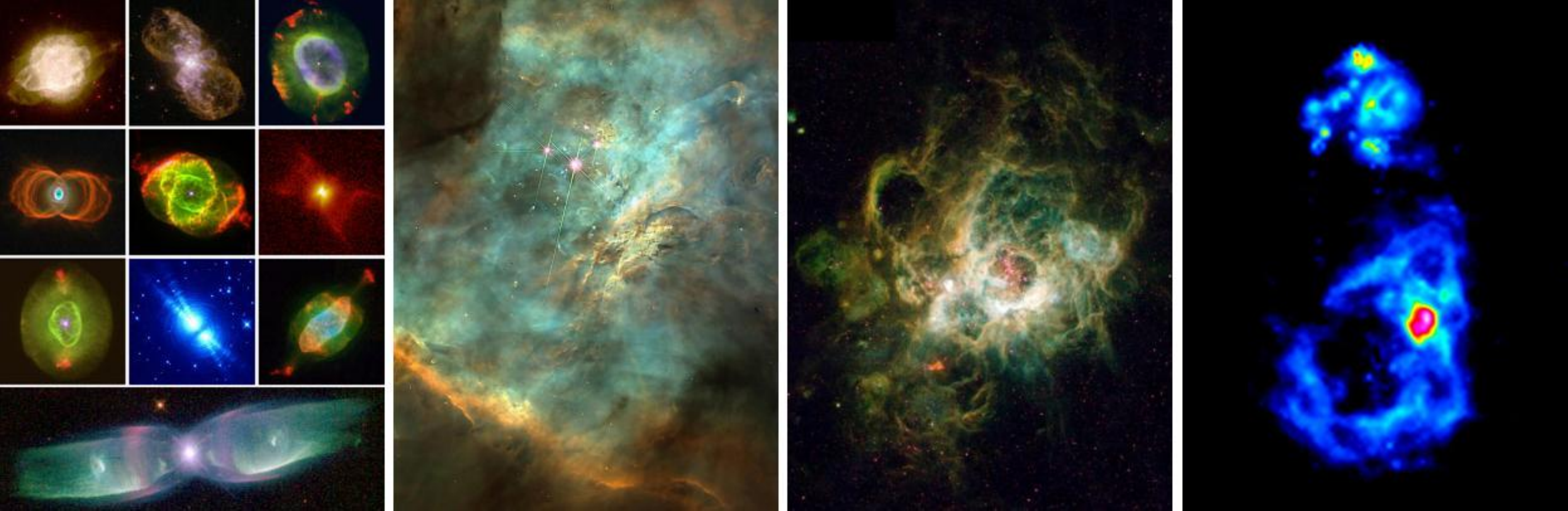}}
\caption{Examples of emission line nebulae. From left to right: a) PNe
(HST images obtained by B. Balick and collaborators; c.f.
http://www.astro.washington.edu/users/balick/WFPC2/index.html); b) The H~{\sc
ii} region M\,42 (the Orion Nebula; Hubble Heritage image obtained by C. R.
O'Dell and S.  K. Wong, c.f.
http://hubblesite.org/newscenter/archive/releases/1995/45); c) NGC\,604, a
giant extragalactic H~{\sc ii} region in the outskirts of the Local Group
spiral galaxy M\,33 (Hubble Heritage image obtained by H. Yang, c.f.
http://heritage.stsci.edu/2003/30/supplemental.html; d) The starburst galaxy
I\,Zw\,18 (based on HST/WFPC2 data obtained by E. Skillman). The linear sizes
of these objects differ vastly, ranging from about a tenth of a parsec
(1\,pc = 3.262\,lyr = $3.086\times 10^{16}$\,m) for PNe and M\,42 to several 
hundred parsecs for NGC\,604 and I\,Zw\,18.} 

\label{fig01} 
\end{figure}

Examples of emission line nebulae include H~{\sc ii} regions, planetary nebulae
(PNe) and the broad and narrow emission line regions found in active galactic
nuclei (Fig.\,\ref{fig01}).  H~{\sc ii} regions are diffuse nebulae found
around newly formed young, massive stars and trace the current status of the
ISM. Giant extragalactic H~{\sc ii} regions, sign posts of massive star
formation activities, are amongst the most prominent features seen in a
gas-rich, star-forming galaxy. In some galaxies, the star forming activities
are so intense that the whole galaxy becomes a giant H~{\sc ii} region. Such
galaxies are called H~{\sc ii} or starburst galaxies and are observable to
large cosmic distances. 

PNe are amongst the most beautiful objects in the sky and, arguably, the queen
of the night. They were given the name by William Herschel (Herschel, 1785)
based on their distinct structures and, for some of them, nearly circular and
overall uniform appearances resembling the greenish disk of a planet. They have
however nothing to do with a planet (but see later) and are in fact expanding
gaseous envelopes expelled by low- and intermediate-mass stars in late
evolutionary stage after the exhaustion of central nuclear fuel at the end of
the asymptotic giant branch (AGB) phase. Owing to their relatively simple
geometric structure, a nearly symmetric shell of gas ionized by a single,
centrally located white dwarf, PNe are ideal cosmic laboratories to study the
atomic and radiative processes governing cosmic low density plasmas. PNe have
played and continue to play a central role in formulating the theory of
photoionized gaseous nebulae.  Representing a pivotal, albeit transient
evolutionary phase of low- and intermediate-mass stars (the overwhelming
majority in a galaxy), PNe play a major role in the galactic ecosystem -- in
the constant enrichment of metals, formation and destruction of molecules and
dust grains and in the recycling of gas in the ISM. Today, studies of PNe have
gone far beyond the objects themselves. PNe are widely used to trace the
kinematics of the host galaxies and the intracluster stellar populations. They
have even been successfully utilized to measure the Hubble constant of the
cosmic expansion.

Recent progress in observational techniques, atomic data and high-performance
computation have enabled reliable measurements and analyses of lines as faint
as one millionth of H$\beta$, including weak optical recombination lines (ORLs)
from abundant heavy elements (C, N, O, Ne and Mg) and collisionally excited
lines (CELs) from rare elements, such as fluorine and s- and r-process
elements. This allows one to address some of the long-standing problems in
nebular astrophysics as well as opening up new windows of opportunity. 

In this contribution, I will briefly review the development of the theory of
photoionized gaseous nebulae, highlighting some of the key events. I will then
present some recent developments of deep spectroscopy of PNe and H~{\sc ii}
regions, concentrating on observations of faint heavy element ORLs. I will show
that there is strong evidence that nebulae contain another previously unknown
component of cold (about 1,000 K), high-metallicity plasma, probably in the
form of H-deficient inclusions embedded in the warm (about 10,000 K) diffuse
nebula of ``normal (i.e. near solar ) composition''. This cold gas emits
essentially all the observed flux of heavy element ORLs, but is too cool to
excite any significant optical or ultraviolet CELs and thus invisible via the
latter. The existence of H-deficient gas in PNe and probably also in H~{\sc ii}
regions, not predicted by the current stellar evolution theory, provides a
natural solution to the long-standing dichotomy between nebular plasma
diagnostics and abundance determinations using CELs on the one hand and ORLs on
the other, a discrepancy that is ubiquitously observed in Galactic and
extragalactic PNe as well as H~{\sc ii} regions.

\section{The founding of the theory of photoionized gaseous nebulae}

The applications of the principle of ``Chemical analysis by observations of
spectra'' expounded by G. Kirchhoff and R. W. Bunsen (1860) by W. Huggins and
W. A. Millar to the analyses of stellar and nebular spectra in 1860s heralded
the rise of astrophysics. In an accompanying paper to their monumental work on
stellar spectra, Huggins and Millar presented their first visual spectroscopic
observations of eight PNe, including the Cat's Eye Nebula NGC\,6543 in Draco
(Huggins and Millar, 1864). Instead of the dark (absorption) Fraunhofer lines
observed in the spectra of the Sun and other stars, they found bright emission
lines in the nebular spectra and concluded that the nebulae must consist of
"enormous masses of luminous gas or vapour". While the Fraunhofer\,F line
(H\,$\beta$ at 4861\,{\AA}) did appear in emission, the two bright, nearby
lines $\lambda\lambda$4959, 5007 were not Fraunhofer lines at all.  The
availability of dry photographic plates (light-sensitive silver bromide salts
held in a gelatine emulsion on the glass), replacing the old wet collodion
plates, made it possible to record long exposures of nebular spectra and more
nebular emission lines were revealed in the blue and ultraviolet (UV)
wavelength regions. In 1882, Huggins successfully photographed a UV spectrum of
the nearest bright H~{\sc ii} region, the Orion Nebula, and detected a strong
emission line near 3728\,{\AA} (Huggins, 1882). 

By the late 1920s, dozens of nebular lines had been detected and their
wavelengths accurately measured (Wright, 1918), yet most of them remained
unidentified and were attributed to some hypothetical element ``nebulium''. The
big breakthrough in understanding nebular spectra came in 1927 from Ira Bowen
who, stimulated by a heuristic speculation by H. N.  Russell that the nebulium
lines must be due to abundant ``atoms of known kinds shining under unfamiliar
conditions'' such as in gas of very low density (Russell, Dugan and Stewart,
1927), identified eight of the strongest nebular lines as being due to the
forbidden transitions from the low excitation meta-stable states of the ground
electron configurations of singly ionized oxygen, nitrogen and doubly ionized
oxygen (Bowen, 1927a, b, c).  For the first time, the physical processes in
gaseous nebulae could be understood. This important discovery paved the way for
future studies of nebular structure and chemical composition.

Great strides forward in nebular spectral observations were made starting in
1910 and through the 1930s, powered by a number of technological inventions,
such as the Schmidt camera (Schmidt, 1932), the high-efficiency blazed
diffraction grating (Wood, 1910) and the image slicer (Bowen, 1938). Deep
exposures revealed faint lines from the refractory elements such as potassium,
calcium, silicon, magnesium and iron, demonstrating that nebulae are
qualitatively made of similar material as stars (Bowen and Wyse, 1939). In the
following three decades, the structures and the underlying physics governing
photoionized gaseous nebulae had been worked out quantitatively, including
photoionization and recombination (Zanstra, 1926, 1927; Str\"{o}mgren, 1939),
heating and cooling (Zanstra, 1931; Baker, Menzel and Aller, 1938; Aller, Baker
and Menzel, 1939; Spitzer, 1948), recombination and collisional excitation
(Baker and Menzel, 1938; Menzel, Aller and Hebb, 1941; Seaton, 1954; Burgess,
1958; Seaton, 1959; Seaton and Osterbrock, 1957), and elemental abundance
determinations (Bowen and Wyse, 1939; Menzel and Aller, 1941; Wyse 1942; Aller
and Menzel, 1945). Origin of PNe, as descendants of red giant stars was also
understood (Shklovsky, 1956).  Photoionization models incorporating all known
physics were constructed (Hjellming, 1966; Goodson, 1967; Harrington, 1968;
Rubin 1968; Flower, 1969) and the models reproduced observations well. The
theory of photoionized gaseous nebulae as it stood in late 1960s was nicely
summarized in Astrophysics of Gaseous Nebulae by Donald Edward Osterbrock
(1974).

\section{CELs and ORLs -- the dichotomy}

\begin{figure}
\centering
\centerline{\includegraphics[width=180mm]{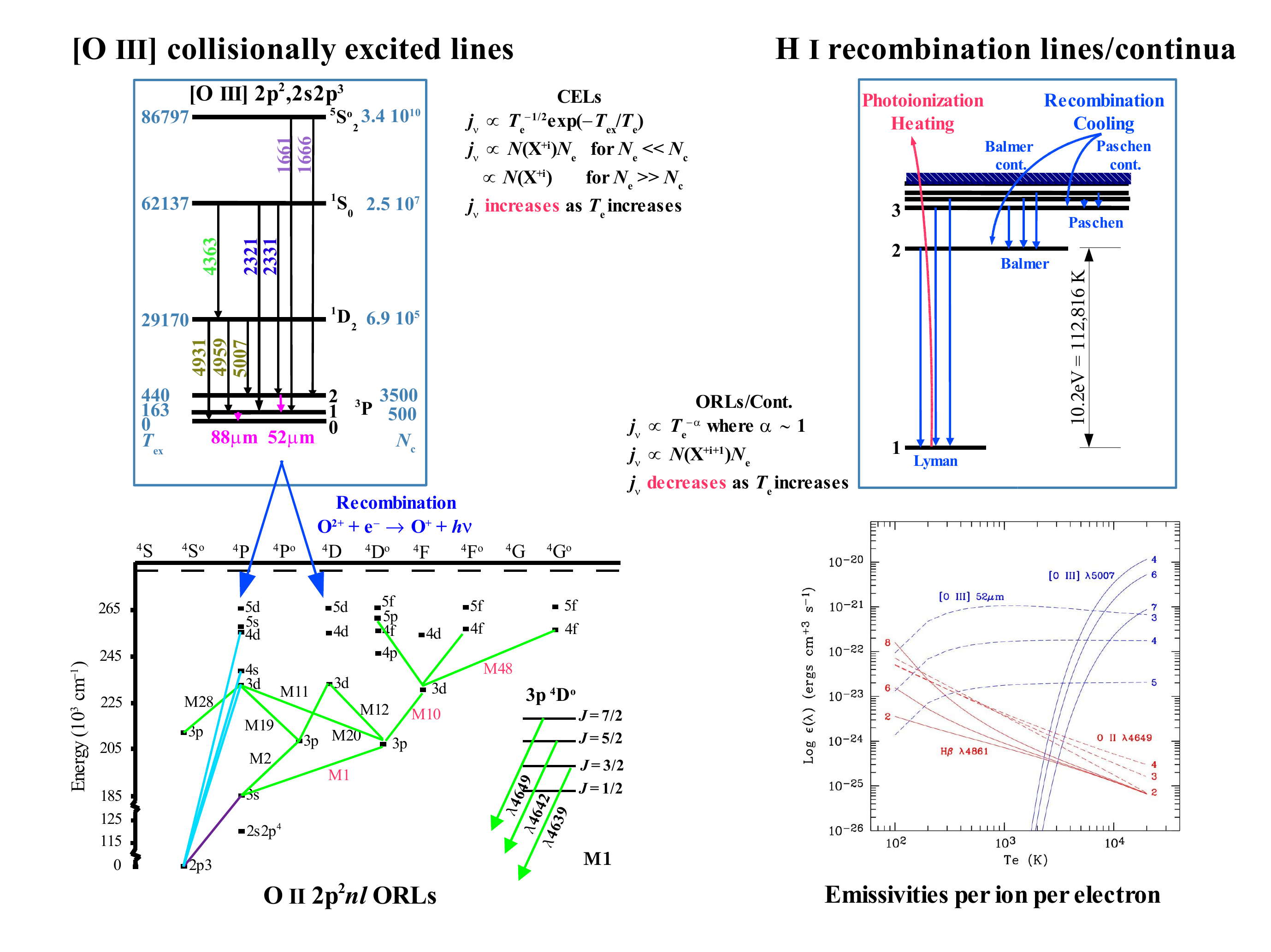}}

\caption{{\em Top-left}: Schematic diagram showing the four lowest spectral
terms of O$^{++}$ formed by the 2p$^2$ ground and 2p$^3$ electron
configurations. The levels are labeled with excitation energy $T_{\rm ex}$ (in
Kelvin) and critical density $N_{\rm c}$ (in cm$^{-3}$); {\em Bottom-left}:
Schematic Grotrian diagram of O~{\sc ii}, illustrating the O~{\sc ii}
recombination line spectrum produced by recombination of O$^{++}$; {\em
Top-right}: Photoionization and recombination of hydrogen. Photoionization
heats the gas whereas recombination cools the gas; {\em Bottom-right}:
Emissivities of the collisionally excited [O~{\sc iii}] $\lambda$5007 optical
forbidden line and 52$\mu$m far-IR fine-structure line, and of the
recombination lines H$\beta$ $\lambda$4861 and O~{\sc ii} $\lambda$4649, as a
function of electron temperature $T_{\rm e}$ for selected electron densities.
The curves are labeled with logarithms of electron densities. Emissivities of
recombination lines depend only weakly on electron density under typical
(low-density) nebular conditions.}

\label{fig02}
\end{figure}

While the theory seemed well established and solid, there were dark clouds
hovering on the horizon. One concerned the measurement and interpretation of
weak nebular emission lines, and the other regarded the possible presence of
significant temperature inhomogeneities in nebulae and their effects on nebular
abundance determinations.

Except for a few lines excited under specific environments, such as the Bowen
fluorescence lines (e.g. O~{\sc iii} $\lambda\lambda$3133, 3341, 3444; Bowen,
1934, 1935), strong lines radiated by photoionized gaseous nebulae fall into
two categories, recombination lines (RLs) and CELs.  Hydrogen and helium ions,
the main baryonic components of an ionized gaseous nebula, capture free
electrons and recombine, followed by cascades to the ground state.  During this
process, a series of RLs are radiated (e.g. H\,$\alpha$ $\lambda$6563,
H\,$\beta$ $\lambda$4861, He~{\sc i} $\lambda\lambda$4472, 5876, 6678 and
He~{\sc ii} $\lambda$4686). The ground electron configuration of multi-electron
ions of heavy elements yields some low excitation energy levels (within a few
eV from the ground state, such as O$^{++}$ 2p$^2$\,$^3$P$^{\rm o}_{0,1,2}$,
$^1$D$^{\rm o}_2$, $^1$S$^{\rm o}_0$), which can be excited by impacts of
thermal electrons, which have typical energies of $\sim 1$\,eV in photoionized
gaseous nebulae of solar composition. Follow-up de-excitation by spontaneous
emission yields the so-called CELs. Quite often, those lines are electron
dipole forbidden transitions, so they are commonly called forbidden lines
(e.g.\, [O~{\sc ii}] $\lambda\lambda$3726,3729, [O~{\sc iii}]
$\lambda\lambda$4959,5007, [N~{\sc ii}] $\lambda\lambda$6548,6584 and
[Ne~{\sc iii}] $\lambda\lambda$3868,3967).  Recombination of free electrons to
bound states of hydrogen and helium ions also yields weak nebular continuum
emission. For example, recombination to the H~{\sc i} $n = 2$ state yields the
near UV nebular continuum Balmer discontinuity at $\lambda < 3646$\,{\AA}.
Recombination of heavy element ions, followed by cascading to the ground state,
also produces RLs. However, for typical cosmic composition, i.e.  approximately
solar, even the most abundant heavy element oxygen has a number density less
than one thousandth of hydrogen. Thus heavy element RLs are much weaker, at the
level of a few thousandth of H$\beta$ or less, and are observable generally
only in the optical. Those lines are therefore often called ORLs. Sample ORLs
from abundant heavy element ions that have been well studied and discussed in
the current contribution include C~{\sc ii} $\lambda$4267, C~{\sc iii}
$\lambda$4649, N~{\sc ii} $\lambda$4041, O~{\sc i} $\lambda$7773, O~{\sc ii}
$\lambda$4089, O~{\sc iii} $\lambda$3265, Ne~{\sc ii} $\lambda$4392 and Mg~{\sc
ii} $\lambda$4481.

Except for IR fine-structure lines arising from ground spectral terms,
emissivities of CELs have an exponential dependence on electron temperature,
$T_{\rm e}$, $\epsilon({\rm X}^{\rm +i}, \lambda) \propto T^{-1/2}_{\rm e}{\rm
exp}(-E_{\rm ex}/kT_{\rm e})$ (the Boltzmann factor), where $E_{\rm ex}$ is
excitation energy of the upper level of transition emitting wavelength
$\lambda$ by ion X$^{\rm +i}$ following collisional excitation by electron
impacts (Fig.\,\ref{fig02}). At low densities, $N_{\rm e} << N_{\rm c}$, i.e.
electron density $N_{\rm e}$ much lower than the upper level's critical density
$N_{\rm c}$ (for any energy level above the ground, a critical density $N_{\rm
c}$ can be defined such that above which collisional de-excitation dominates
over spontaneous radiative decay in de-populating the level; c.f.  Osterbrock
\& Ferland, 2005), we have $\epsilon({\rm X}^{\rm +i}, \lambda) \propto N({\rm
X}^{\rm +i})N_{\rm e}$, where $N({\rm X}^{\rm +i})$ is number density of ion
X$^{\rm +i}$.  At high densities, $N_{\rm e} >> N_{\rm c}$, emission of CELs
are suppressed by collisional de-excitation and $\epsilon({\rm X}^{\rm +i},
\lambda) \propto N({\rm X}^{\rm +i})$. Unlike CELs, emissivities of RLs {\em
increase} with {\em decreasing}\ $T_{\rm e}$ by a power law. Under typical
nebular conditions ($N_{\rm e} << 10^8\,{\rm cm}^{-3}$), $\epsilon({\rm X}^{\rm
+i}, \lambda) \propto T_{\rm e}^{-\alpha}N({\rm X}^{\rm +i+1})N_{\rm e}$, where
$\alpha \sim 1$ and $N({\rm X}^{\rm +i+1})$ is number density of the {\em
recombining ion}\ X$^{\rm +i+1}$.

In a ground breaking work, Arthur B. Wyse published very deep spectra that he
obtained with I. S. Bowen using the Lick 36-inch reflector for a number of
bright PNe (Wyse, 1942). About 270 spectral lines were detected. Many were weak
permitted transitions from abundant C, N and O ions. In the Saturn Nebula
NGC~7009 alone, ``... more than twenty lines or blends of O~{\sc ii} have been
observed, free of blending with lines of other elements. There are good grounds
for the assumption, previously made, that these lines originate in electron
captures by O~{\sc iii} ions, just as the Balmer lines originate in electron
captures by H~{\sc ii} ions, and that therefore the relative intensities of the
oxygen and hydrogen lines give a measure of the rates of recombination, and
therefore of the relative abundance, of the two kinds of ions .... The
important thing is that for this nebula a sufficient number of oxygen lines has
been observed to eliminate all doubt of the correctness of their
identifications, even though they are very faint; and also that they permit the
direct estimate of the relative abundance of H~{\sc ii} and O~{\sc iii} ions,
..., by a method that is relatively independent of assumptions regarding
electron density and velocity distribution or their variations throughout the
nebula.''

From the O~{\sc ii} ORL strengths, Wyse deduced O/H abundances which were much
higher than the values obtained by Menzel and Aller (1941) from analyses of the
[O~{\sc iii}] forbidden lines. He concluded ``The discrepancy, then, between
the relative abundance found by Menzel and Aller, on the one hand, and from the
recombination spectra, on the other, is of the order of 50 for NGC\,7027 and of
500 for NGC\,7009.'' Shortly after the publication of this prophetic article,
Wyse was called to serve in the War and died tragically on duty the night of
June 8, 1942 in a disastrous accident over the Atlantic Ocean, off the New
Jersey coast, at the age of 17 days shy of 33.

As stressed by Wyse, ionic abundances deduced from intensities of heavy element
ORLs relative to H$\beta$, a method based on comparing lines of like to like,
have the advantage that they are almost independent of the nebular thermal and
density structures. In contrast, ionic abundances deduced from the intensity
ratio of the collisionally excited, much stronger [O~{\sc iii}]
$\lambda\lambda$4959,5007 forbidden lines relative to H$\beta$, have an
exponential dependence on the adopted nebular electron temperature.
Theoretically, ionic abundances deduced from heavy element ORLs should thus be
more reliable, provided the lines can be measured accurately. Unfortunately,
latter development showed that accurate flux measurements for faint nebular
emission lines were NOT possible after all with the technique available then,
i.e.  spectrophotography, due to the non-linearity of photographic plates. Via
detailed comparisons between the observed fluxes of H~{\sc i} and He~{\sc ii}
RLs and continua and those predicted by the recombination theory, it became
clear that spectrophotographic observations systematically overestimated
intensities of faint lines, by as much as over a factor of ten (Seaton, 1960;
Kaler, 1966; Miller, 1971; Miller and Mathews, 1972). That spectrophotographic
measurements of faint lines cannot be trusted seemed to be further supported by
work in the 1980s that contrasted C$^{++}$/H$^+$ ionic abundances deduced from
the collisionally excited C~{\sc iii}] $\lambda\lambda$1907,1909
intercombination lines and from the faint C~{\sc ii} $\lambda$4267 ORL (c.f.
Barker, 1991 and references therein), suggesting that the intensity of the
faint C~{\sc ii} $\lambda$4267 line had either not been interpreted correctly
or been grossly overestimated (Rola and Stasi\'{n}ska, 1994).
 
In another worrisome development, Manual Peimbert showed that if nebulae are
non-isothermal and have (localized and random) temperature fluctuations, then
$T_{\rm e}$ deduced from the [O~{\sc iii}] nebular to auroral line intensity
ratio ($\lambda$4959 + $\lambda$5007)/$\lambda$4363 will overestimate the
average emission temperature of the $\lambda\lambda$4959,5007 nebular lines and
of H$\beta$ at 4861\,{\AA}, leading to underestimated O$^{++}$/H$^+$ ionic
abundance ratio deduced from the ($\lambda$4959 + $\lambda$5007)/H$\beta$ ratio
(Peimbert, 1967).  He presented evidence pointing to the presence of large
temperature fluctuations in PNe and H~{\sc ii} regions by finding that $T_{\rm
e}$'s derived from the Balmer jump, $T_{\rm e}$(BJ), are systematically lower
than those derived from the [O~{\sc iii}] forbidden line ratio, $T_{\rm e}$([O
III]) (Peimbert 1967; Peimbert, 1971). Given the weakness of nebular continuum
emission, measuring the Balmer jump accurately was no easy task and his results
were disputed (Barker, 1978). Theoretically, while some systematic spatial
temperature variations undoubtedly occur within a nebula, due to changes of
ionization structure and cooling rates as a function of position induced by
varying ionization radiation field and density distribution, no mechanisms are
known capable of generating large, localized temperature fluctuations,
certainly nothing of the magnitude implied by Peimbert's measurements, which
yield a typical value of 0.055 for the temperature fluctuation parameter,
$t^2$, or fluctuations of an amplitude of 23 per cent.

The advent of linear, high quantum efficiency, large dynamic range and large
format CCDs in the 1980s made it possible for the first time to obtain reliable
measurements of faint emission lines for bright nebulae. Meanwhile, the
completion of the Opacity Project (Seaton, 1987; Cunto et al., 1993) has
allowed the atomic data necessary to analyze those spectral features,
specifically their effective recombination coefficients, to be calculated with
high accuracy. Liu and Danziger (1993) obtained CCD measurements of the Balmer
jump for a sample of PNe and found that $T_{\rm e}$(BJ) is indeed
systematically lower than $T_{\rm e}$([O~{\sc iii}]) obtained for the same
object. They deduced that on average, $t^2 = 0.035$, smaller than that found
earlier by Peimbert (1971) but still significant, enough to cause the
O$^{++}$/H$^+$ abundance ratio derived from the $\lambda\lambda$4959,5007
forbidden lines to be underestimated by a factor of two. 

Liu et al. (1995) presented high quality IPCS and CCD optical spectrophotometry
for the legendary Saturn Nebula NGC\,7009 as well as new effective
recombination coefficients for O~{\sc ii} ORLs. Nearly a hundred O~{\sc ii}
ORLs were measured, yielding an O$^{++}$/H$^+$ ionic abundance that is
consistently higher, by a factor of $\sim 4.7$, than the value deduced from the
strong [O~{\sc iii}] forbidden lines $\lambda\lambda$4959,5007. The close
agreement of results deduced from a large number of O~{\sc ii} ORLs from a
variety of multiplets of different multiplicities, parities and electron
configurations, vindicates the reliability of the recombination theory and
rules out measurement uncertainties\footnote{Mathis and Liu (1999) analyzed the
observed relative intensities of the [O~{\sc iii}] $\lambda\lambda$4959,5007
and the much fainter $\lambda$4931 nebular lines and demonstrated that accurate
measurements have been achieved over a dynamic range of 10,000.} or other
effects, such as reddening corrections, line blending or contamination of ORLs
by other excitation mechanisms (e.g. fluorescence or charge transfer reactions)
as the cause of the large discrepancy between the ORL and CEL abundances.
Further analysis of the carbon, nitrogen recombination spectra and of the neon
recombination spectrum (Luo et al., 2001) show that in NGC\,7009, abundances of
these elements derived from ORLs are all higher than the corresponding CEL
values by approximately a factor of 5. If one defines an abundance discrepancy
factor (adf) as the ratio of ionic abundances X$^{\rm i+}$/H$^+$ deduced from
ORLs and from CELs, then in NGC\,7009 ${\rm adf} \sim 5$ for all the four
abundant second-row elements of the periodic table: C, N, O and Ne. In
NGC\,7009, the Balmer discontinuity of the hydrogen recombination spectrum
yields $T_{\rm e}$(BJ) $= 8,100$\,K, about 2,000\,K lower than forbidden line
temperature $T_{\rm e}$([O~{\sc iii}]) $= 10,100$\,K (Liu et al., 1995). The
difference yields a Peimbert's $t^2$ value of 0.04, or temperature fluctuations
of an amplitude of 20 per cent.

\section{Deep spectroscopic surveys of ORLs}

Is the dichotomy observed in NGC\,7009 between ORLs and CELs for nebular plasma
diagnostics and abundance determinations ubiquitous amongst emission line
nebulae? What is the range and distribution of adf's for individual elements?
Are there any correlations between the adf and other nebular properties
(morphology, density, temperature, abundance, and age etc.) or properties of
the ionizing source? What are the physical causes of the dichotomy? To address
those issues, several deep ORL spectroscopic surveys of faint heavy element
ORLs have been carried out. So far over a hundred Galactic PNe have been
surveyed, plus dozens of Galactic and extragalactic H~{\sc ii} regions (e.g.
Esteban et al., 2002; Tsamis et al., 2003a,b, 2004; Liu et al., 2004a,b; Wesson
et al., 2005; Wang and Liu, 2007; For a complete list of references, please
refer to a recent review by Liu, 2006a).  Detailed comparisons contrasting ORL
and CEL analyses show that (Fig.\,\ref{fig03}; c.f. also Liu, 2006a):

\begin{figure}
\centering
\centerline{\includegraphics[width=160mm]{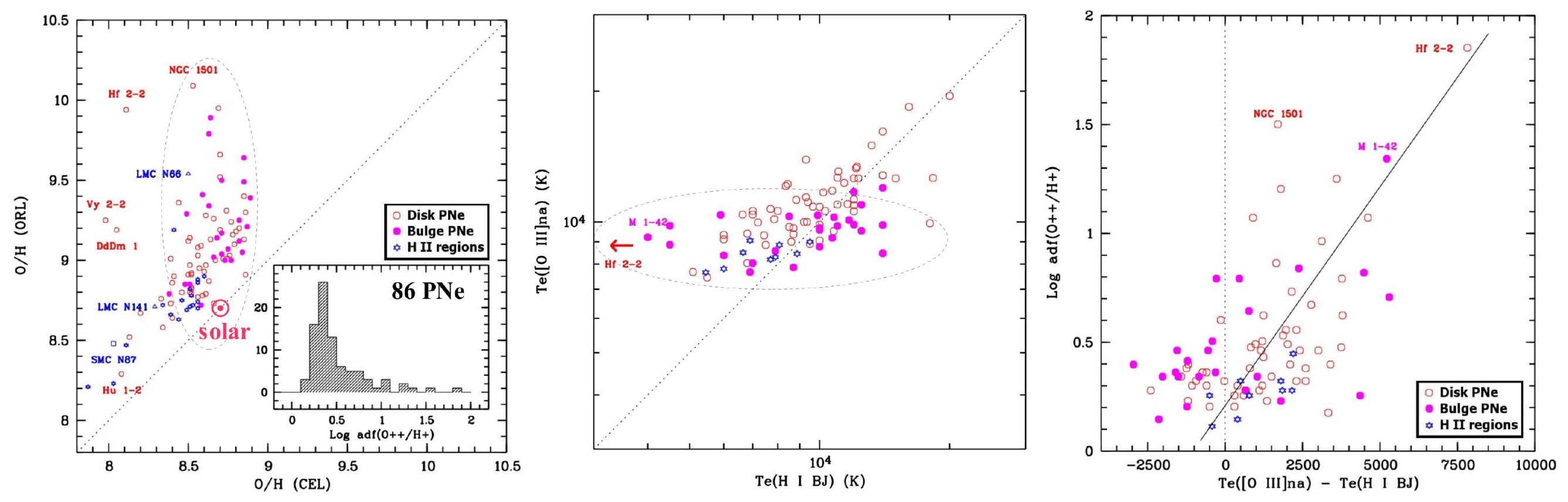}}

\caption{{\em Left}: O/H abundances deduced from ORLs plotted against those
derived from CELs. The diagonal dotted line denotes $x = y$. The insert is a
histogram of adf(O$^{++}$/H$^+$). In the scenario of single composition nebulae
(c.f. next section for an alternative interpretation), uncertainties in O/H
abundances, caused by observational and interpretive errors (e.g. atomic data)
are expected to be typically less than 0.05 and 0.1\,dex, respectively, for CEL
and ORL results; {\em Middle}: $T_{\rm e}$([O~{\sc iii}]) versus $T_{\rm
e}$(BJ). Typical uncertainties of $T_{\rm e}$([O~{\sc iii}]) and $T_{\rm
e}$(BJ) are 5 and 10 per cent, respectively. The diagonal dotted line denotes
$x = y$. With $T_{\rm e}({\rm BJ}) = 900$\,K and $T_{\rm e}$([O~{\sc iii}]) $ =
8820$\,K, Hf\,2-2 falls off the left boundary of the plot; {\em Right}: ${\rm
log}$\,adf(O$^{++}$/H$^+$) plotted against $T_{\rm e}$([O~{\sc iii}]) $-$
$T_{\rm e}$(BJ). The solid line denotes a linear fit obtained by Liu et al.
(2001b), prior to the discovery of the very large adf in Hf\,2-2.  (Adapted
from Liu, 2006a)}

\label{fig03}
\end{figure}

\begin{figure}
\centering
\centerline{\includegraphics[width=160mm]{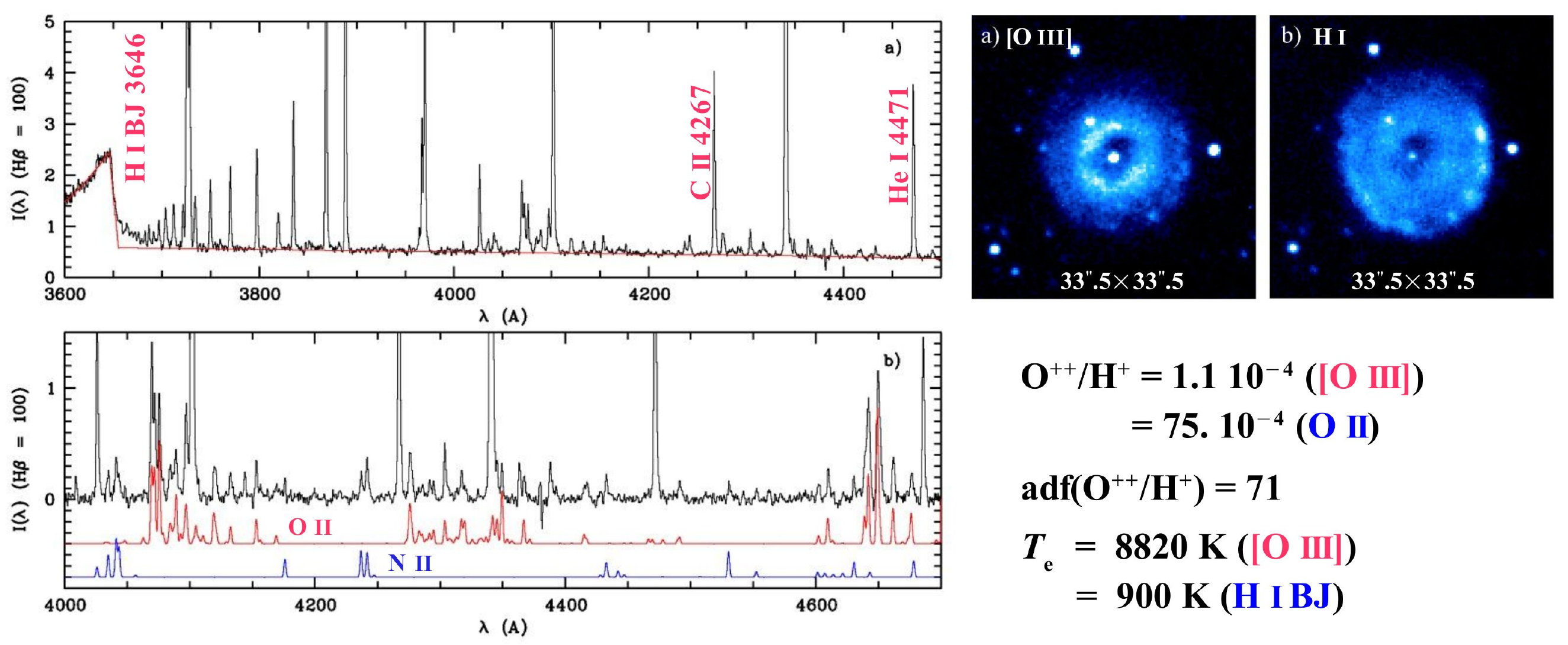}}

\caption{{\em Left}: The optical spectrum of Hf\,2-2, which shows a record
adf(O$^{++}$/H$^+$) $ = 71$ and an extremely low $T_{\rm e}$(BJ) of 900\,K. In
the lower panel, also shown are two synthetic recombination line spectra of
O~{\sc ii} and N~{\sc ii}, respectively. {\em Right}: Monochromatic images of
Hf\,2-2 in the light of [O~{\sc iii}] $\lambda$5007 and H$\alpha$
$\lambda$6563, respectively. (Adapted from Liu et al., 2006)}

\label{fig04}
\end{figure}

1) Ionic abundances deduced from ORLs are {\em always} higher than CEL values,
i.e. adf $\geq 1$. Adf peaks at 0.35\,dex but with a tail extending to much
higher values. About 20 and 10 per cent of nebulae exhibit adf's higher than 5
and 10, respectively. For example, in the bright Galactic disk PN NGC\,6153,
adf = 9.2 (Liu et al., 2000), whereas in the bulge PN M\,1-42, adf = 22 (Liu et
al., 2001b). In the most extreme object discovered so far, Hf\,2-2,
adf(O$^{++}$/H$^+$) reaches a record value of 71 (Fig.\,\ref{fig04}); 

2) While adf varies from object to object, for a given nebula, C, N, O and Ne
all exhibit comparable adf's (thus both CEL and ORL analyses yield compatible
abundance ratios, such as C/O, N/O and Ne/O, provided lines of the same type,
ORLs or CELs, are used for both elements involved in the ratio). Objects
showing large adf's also tend to have high helium (ORL) abundances. However,
magnesium, the only third-row element which has been analyzed using an ORL,
shows no enhancement, even in high-adf objects (Barlow et al., 2003);
 
3) Excluding metal-poor nebulae in the Galactic halo and in the Large and Small
Magellanic Clouds (LMC and SMC, respectively), oxygen abundances deduced from
CELs for Galactic H~{\sc ii} regions and PNe fall in a narrow range compatible
with the solar value. In contrast, ORLs yield much higher abundances, more than
ten times solar in some cases. 

4) Similarly, while the [O~{\sc iii}] forbidden line ratio yields values of
$T_{\rm e}$ in a narrow range around 10,000\,K, as one expects for a
photoionized gaseous nebula of solar composition, the Balmer discontinuity
yields some very low temperatures, down to below 1,000\,K. In fact, the
discrepancies in temperature and abundance determinations, using ORLs/continua
on the one hand and CELs on the other, seem to be correlated -- objects showing
large adf's also exhibit very low $T_{\rm e}$'s;

5) Large, old PNe of low surface brightness tend to show higher adf's.  In
addition, spatially resolved analyses of a limited number of bright, extended
nebulae of large adf's show that ORL abundances increase towards the nebular
center, leading to higher adf's near the center.  
 
\section{Evidence of cold, H-deficient inclusions}

What causes the ubiquitous, often alarmingly large, discrepancies between the
ORL and CEL plasma diagnostics and abundance determinations?  Does the
dichotomy imply that there are fundamental flaws in our understanding of the
nebular thermal structure, or that we do not even understand basic processes
such as the recombination of hydrogenic ions?

Can it be temperature fluctuations as originally postulated by Peimbert (1967)?
This conjecture implicitly assumes that the higher abundances yielded by ORLs
represent the true nebular composition as they are insensitive to temperature
and temperature fluctuations. The discovery of nebulae exhibiting extreme
values of adf and the abnormally high metal abundances implied by the observed
strengths of ORLs casts however serious doubt on this paradigm. Given that PNe
are descendants of low- and intermediate-mass stars, the very high ORL oxygen
abundances recorded in objects of extreme adf's, if real and representative of
the whole nebula, are extremely difficult to understand in the current theory
of stellar evolution and nucleosynthesis. 

\begin{figure}
\centering
\centerline{\includegraphics[width=160mm]{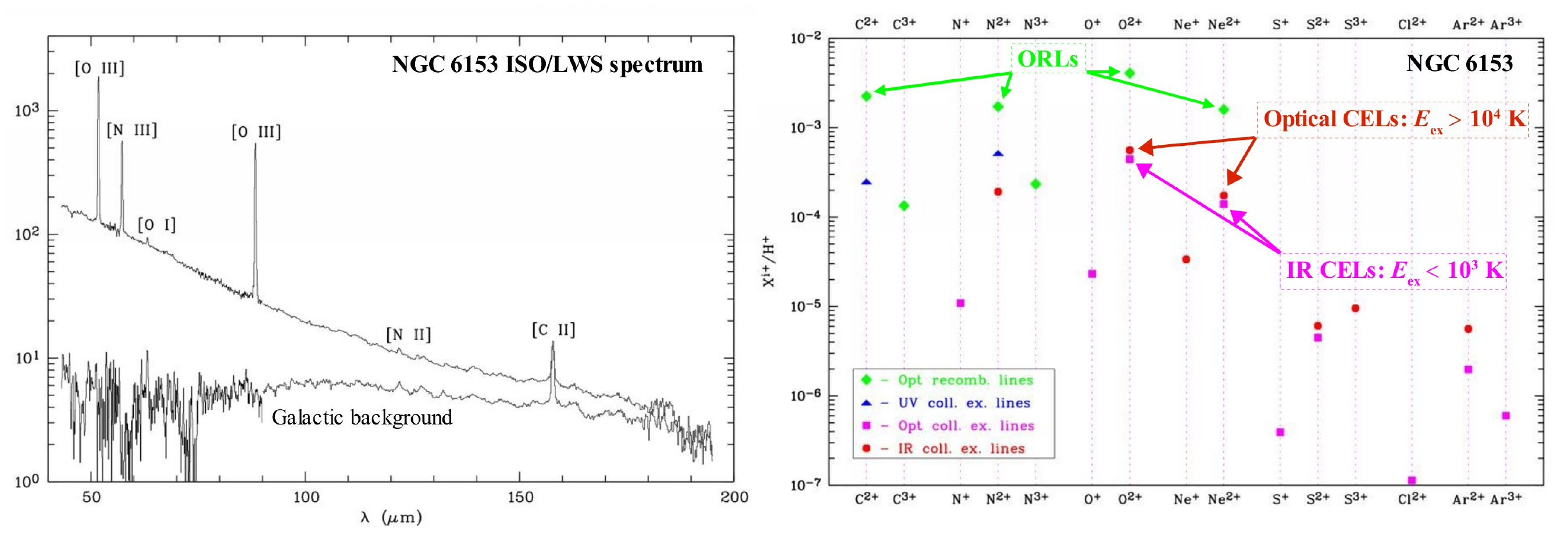}}

\caption{{\em Left}: The far-IR spectrum of NGC\,6153; {\em Right}: Comparisons
of ionic abundances of NGC\,6153 deduced from ORLs, and from UV, optical and IR
CELs. (Adapted from Liu et al., 2000)}

\label{fig05}
\end{figure}

Strong evidence against temperature fluctuations as the cause of the dichotomy
between ORLs and CELs is provided by Infrared Space Observatory (ISO)
measurements of mid- and far-IR fine-structure lines, such as the [Ne~{\sc
iii}] 15.5-$\mu$m and [O~{\sc iii}] 52- and 88-$\mu$m (Liu et al., 2001a).
Although collisionally excited, IR fine-structure lines have, unlike their
optical and UV counterparts, excitation energies less than $\sim 1,000$\,K
(Fig.\,\ref{fig02}), and are therefore insensitive to temperature or
temperature fluctuations. If temperature fluctuations are indeed at work,
leading to an overestimated $T_{\rm e}$([O~{\sc iii}]) and consequently an
underestimated $\lambda\lambda$4959,5007 O$^{++}$/H$^+$ abundance, then one
expects a higher abundance from the 52- and 88-$\mu$m fine-structure lines
comparable to the value yielded by O~{\sc ii} ORLs. ISO measurements and
subsequent analyses reveal, however, otherwise. Fig.\,\ref{fig05} shows that in
the case of NGC\,6153, all CELs -- UV, optical and IR, regardless of their
excitation energy and critical density, yield ionic abundances that are
consistently a factor of $\sim 10$ lower than ORLs.

To account for the multi-waveband observations of NGC\,6153, Liu et al. (2000)
postulated that the nebula contains a previously unknown component of
high-metallicity gas, presumably in the form of H-deficient inclusions embedded
in the diffuse nebula of ``normal'' (i.e. about solar) composition. Due to the
efficient cooling of abundant metals, this H-deficient gas has an electron
temperature of only $\sim 1,000$\,K, too low to excite any optical or UV CELs
(thus invisible via the latter). Yet, the high metallicity combined with a very
low electron temperature make those H-deficient inclusions powerful emitters of
heavy element ORLs. In this picture, ORLs and CELs yield discrepant electron
temperatures and ionic abundances because they probe two different gas
components that co-exist in the same nebula but have vastly different physical
and chemical characteristics. Empirical analysis of Liu et al. (2000) as well
as follow-up 1D photoionization modeling (P\'{e}quignot et al., 2002) shows
that a small amount of H-deficient material, about one Jupiter mass, is
sufficient to account for the strengths of heavy element ORLs observed in
NGC\,6153.

The increasingly lower values of Balmer jump temperature $T_{\rm e}({\rm BJ})$
found for nebulae of larger and larger adf's: 6,000\,K in NGC\,6153 (adf =
9.2), 4,000\,K in M\,1-42 (adf = 22) and 900\,K in Hf\,2-2 (adf = 71), provide
the smoking gun evidence that nebulae contain two regimes of vastly different
physical properties. Further evidence is provided by careful analyses of the
He~{\sc i} and heavy element recombination line spectra, which show that the
average emission temperatures of the He~{\sc i} and O~{\sc ii} ORLs are even
lower than indicated by the H~{\sc i} Balmer discontinuity (Liu 2003).  In
general, it is found that, $T_{\rm e}$(O~{\sc ii}) $\leq$ $T_{\rm e}$(He~{\sc
i}) $\leq$ $T_{\rm e}$(BJ) $\leq$ $T_{\rm e}$([O~{\sc iii}])
(Table\,\ref{tab01}; c.f. Liu, 2006a and references therein), as one expects in
the dual-abundance scenario proposed by Liu et al.  (2000).

\begin{table}
\begin{center}
\caption{Comparison of $T_{\rm e}$'s deduced from CELs and from ORLs/continua$^a$}
\label{tab01}
\begin{tabular}{lccccc}
\noalign{\hrule}
\noalign{\vskip3pt}
    Nebula & adf(O$^{++}$/H$^+$) & $T_{\rm e}$([O~{\sc iii}]) & $T_{\rm e}$(BJ) & $T_{\rm e}$(He~{\sc i}) & $T_{\rm e}$(O~{\sc ii}) \\
           &                     &                        (K) &        (K)      &                     (K) &                     (K) \\
\noalign{\vskip3pt}
\noalign{\hrule}
\noalign{\vskip3pt}
NGC 7009  & 4.7 & 9980 & 7200 & 5040 &  420 \\
H 1-41    & 5.1 & 9800 & 4500 & 2930 & $<$288 \\
NGC 2440  & 5.4 &16150 &14000 &      & $<$288 \\
Vy 1-2    & 6.2 &10400 & 6630 & 4430 & 3450 \\
IC 4699   & 6.2 &11720 &12000 & 2460 & $<$288 \\
NGC 6439  & 6.2 &10360 & 9900 & 4900 &  851 \\
M 3-33    & 6.6 &10380 & 5900 & 5020 & 1465 \\
M 2-36    & 6.9 & 8380 & 6000 & 2790 &  520 \\
IC 2003   & 7.3 &12650 &11000 & 5600 & $<$288 \\
NGC 6153  & 9.2 & 9120 & 6000 & 3350 &  350 \\
DdDm 1    &11.8 &12300 &11400 & 3500 &      \\
Vy 2-2    &11.8 &13910 & 9300 & 1890 & 1260 \\
NGC 2022  &16.0 &15000 &13200 &15900 & $<$288 \\
NGC 40    &17.8 &10600 & 7000 &10240 &      \\
M 1-42    &22.0 & 9220 & 4000 & 2260 & $<$288 \\
NGC 1501  &31.7 &11100 & 9400 &      &      \\
Hf 2-2    &71.2 & 8820 & 900 &  940 &  630 \\\hline
\end{tabular}
\begin{description}
$^a$ Adapted from Liu (2006a). $T_{\rm e}$(He~{\sc i})'s were derived
from the He~{\sc i} $\lambda$5876/$\lambda$4472 and $\lambda$6678/$\lambda$4472
line ratios and were typically accurate to 20 per cent (c.f. Zhang et al., 
2005). Values of $T_{\rm e}$(O~{\sc ii}) were deduced from the O~{\sc ii}
$\lambda$4089/$\lambda$4649 line ratio and had typical uncertainties of 30
per cent.
\end{description}
\end{center}
\end{table}

Detailed 3D photoionization models of NGC\,6153 with and without H-deficient
inclusions have been constructed by Yuan et al. [2010, in preparation; c.f.
Figs.\,\ref{fig06}a) and b)] using MOCASSIN, a Monte Carlo photoionization code
capable of dealing with nebulae of arbitrary geometry and composition (Ercolano
et al.  2003a). In their models, the main nebula was modeled with a chemically
homogeneous ellipsoid of ``normal composition'' (i.e. about solar as yielded by
the CEL analysis). To mimic the bipolar shape of the nebula, the density in the
ellipsoid was allowed to decrease from the equator to the poles.  In addition,
to reproduce the strengths of low-ionization lines, such as [C~{\sc i}]
$\lambda\lambda$9824,9850, [N~{\sc i}] $\lambda\lambda$5198,5200, [N~{\sc ii}]
$\lambda\lambda$6548,6584, [O~{\sc i}] $\lambda\lambda$6300,6363 and [O~{\sc
ii}] $\lambda\lambda$3727,3729, an equatorial ring of the same chemical
composition but of a higher density was added. The presence of a high-density
torus is supported by a high resolution spectrum obtained with the Manchester
Echelle Spectrograph mounted on the Anglo-Australian Telescope.  The spectrum,
centered on H$\alpha$ and the [N~{\sc ii}] $\lambda\lambda$6548,6584 lines and
obtained with a long-slit oriented in PA = 123\,deg and through the central
star, revealed two high velocity emission spots at the positions of the bright
shell, one on each side of the central star. The spots are particularly bright
in [N~{\sc ii}] and have respectively blue- and red-shifted velocities relative
to the H$\alpha$ emission from the nearby nebular shell. The model fits the CEL
measurements well except for a few lines [c.f. the top panel of
Fig.\,\ref{fig06}b)]. The model still underestimates the [C~{\sc i}]
$\lambda$9850 line by a factor of three even after the introduction of the
high-density torus. The model also underestimates, by about a factor of two,
the fluxes of the N~{\sc iii}] $\lambda$1744 and [Ar~{\sc iii}] 8.9- and
21.8-$\mu$m lines. The latter three lines were however only marginally
detected, by the IUE and ISO, respectively, and the discrepancies may well be
due to measurement errors. The largest discrepancy is found for the [Ne~{\sc
ii}] 12.8-$\mu$m line where the model underestimates the observation by more
than a factor of five. 

\begin{figure}
\centering
\centerline{\includegraphics[width=135mm]{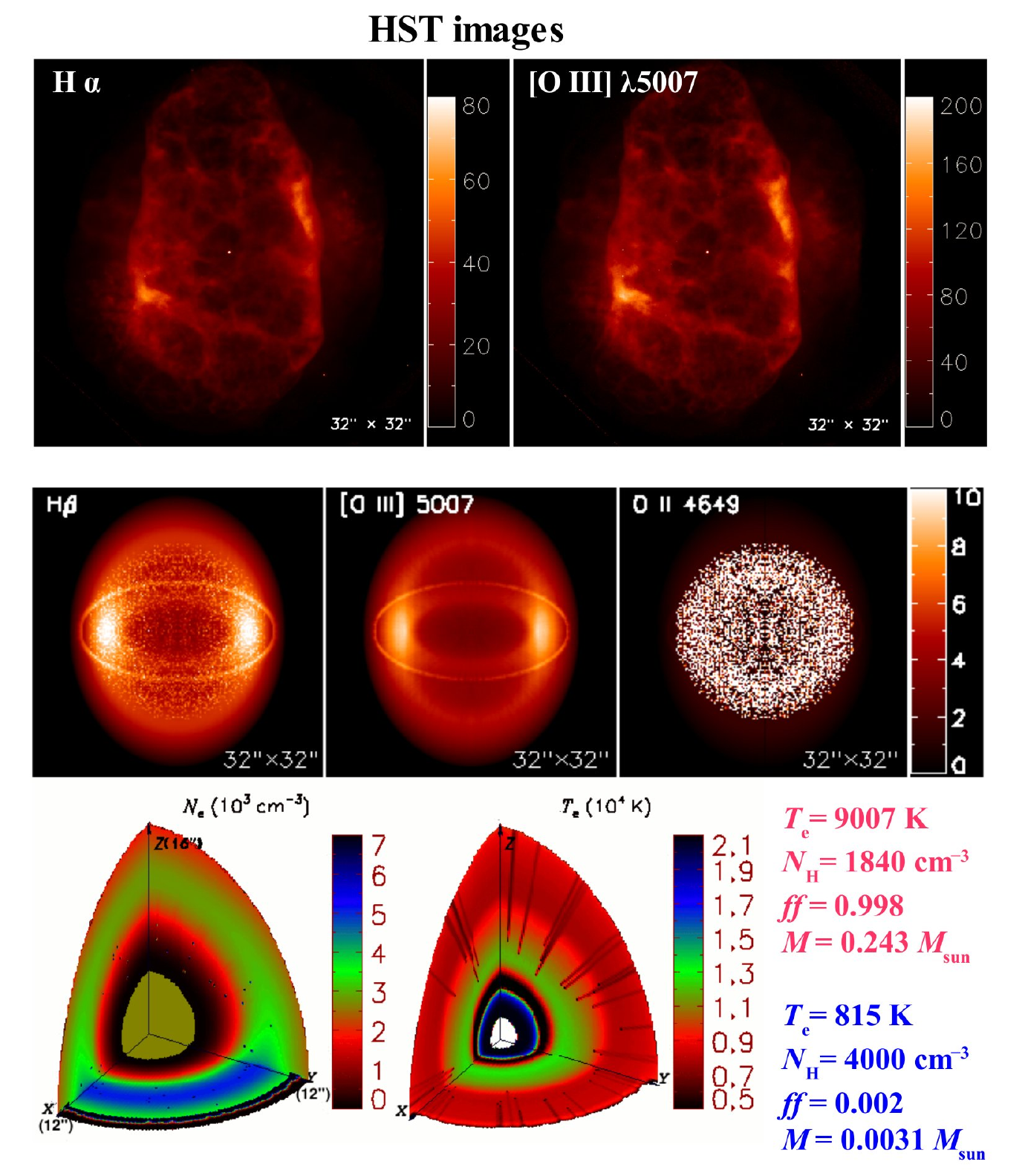}}

\caption{a) From top to bottom: HST monochromatic images of NGC\,6153
in H$\alpha$ and [O~{\sc iii}] $\lambda$5007 (north is up and east to the
left); monochromatic images predicted by the best-fit model incorporating
H-deficient inclusions; and the model distributions of electron density and
temperature.}

\label{fig06}
\end{figure}

\setcounter{figure}{5}

\begin{figure}
\centering
\centerline{\includegraphics[width=135mm]{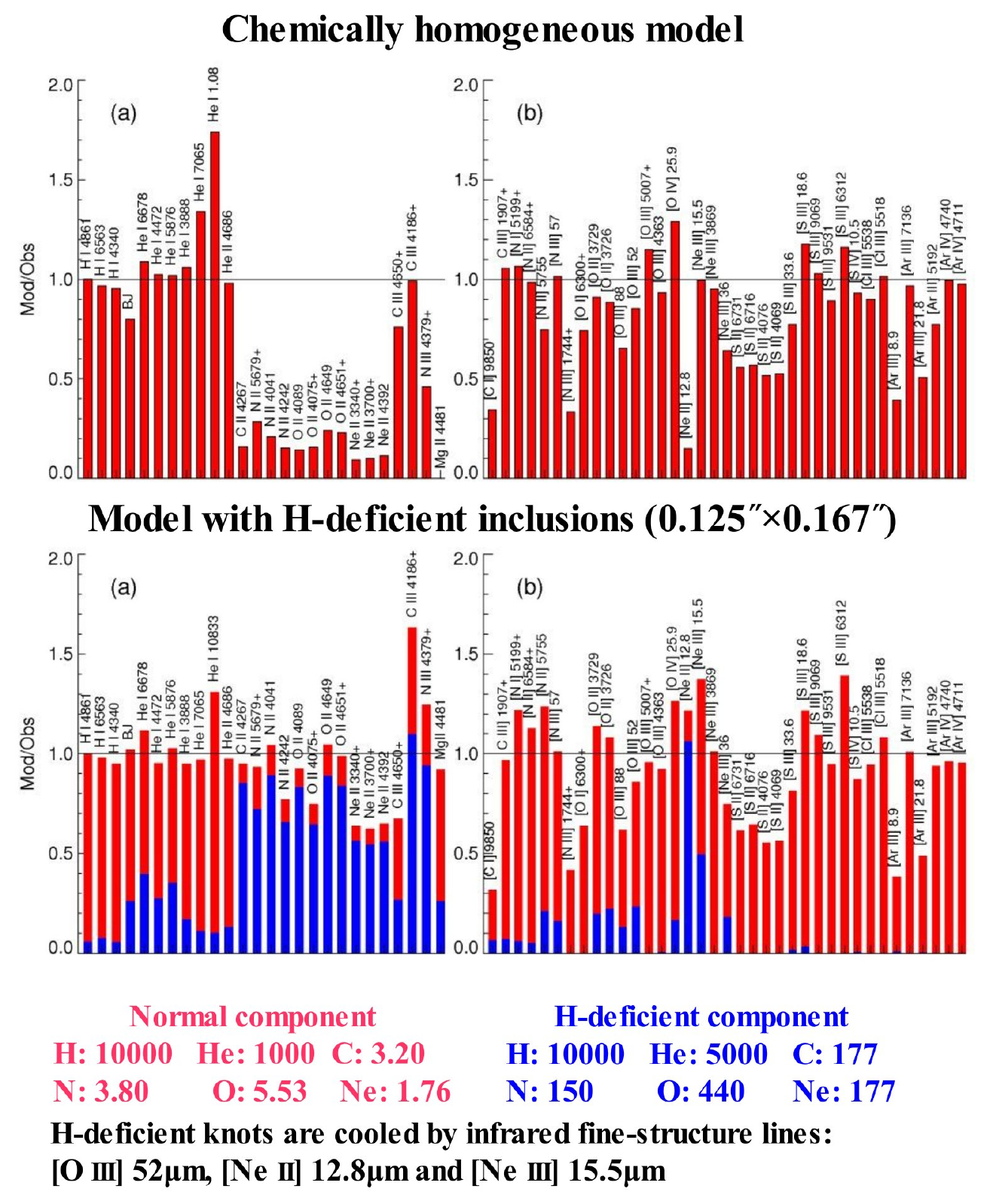}}

\caption{-- {\em continued.}\ b) Comparison of model line intensities with
observations for the best-fit chemically homogeneous model (top) and model
containing H-deficient inclusions (bottom). In the bottom panel, blue and red
color bars denote, respectively, contributions of line fluxes from the cold,
H-deficient inclusions and from the hot, diffuse gas of ``normal
composition''.}

\label{fig06}
\end{figure}

The chemically homogeneous model however fails to reproduce the strengths of
all heavy element ORLs by a factor of ten. To account for them, metal-rich
inclusions were added, in the form of clumps of H-deficient material. The
clumps were distributed spherically symmetrically with a radial number density
profile that reproduced the O~{\sc ii} and C~{\sc ii} ORL surface brightness
distributions deduced from the long-slit observations (Liu et al., 2000). Given
that the H-deficient inclusions have not been resolved, even with the HST/STIS
spectroscopy with a slit width of 0.2\,arcsec (Yuan et al., 2010), they must
have dimensions smaller than $\sim 0.2$\,arcsec, or 300\,AU at the assumed
distance of 1.5\,kpc of NGC\,6153. The smallest size of inclusions that can be
modeled with the MOCASSIN was limited by the memory available per processor of
the computer employed. For NGC\,6153, a grid of $48^3$ cells, each of
1/4\,arcsec by 1/3\,arcsec, was used to model 1/8 of the nebula using a SGI
Altix 330 cluster of 8 dual-core processors and 8\,Gb memory per processor.
Models that doubled the resolution were also constructed using the NASA/Ames
Columbia Supercomputer. Results from one of these high resolution models are
shown in Fig.\,\ref{fig06}a) and b).  The best-fit dual-abundance model
incorporating H-deficient inclusions matches all available observations
reasonably well and successfully reproduces strengths of heavy element ORLs.
With a total mass of only one Jupiter mass, or just over 1\% that of the whole
nebula, a helium abundance 5 times and CNONe abundances 40 -- 100 times higher
than the main diffuse nebula of ``normal composition'', the H-deficient
inclusions account for approximately, 5, 35 and 100\% of the observed fluxes of
H~{\sc i}, He~{\sc i} and heavy element ORLs, respectively, but produce
essentially nil optical and UV CEL emission. Cooled to about 800\,K by IR
fine-structure lines of heavy elemental ions, notably the [Ne~{\sc ii}]
12.8-$\mu$m, [Ne~{\sc iii}] 15.5-$\mu$m and [O~{\sc iii}] 52-$\mu$m lines, the
inclusions are important contributors of those lines, so for the H~{\sc i}
recombination continuum Balmer discontinuity (about 30\%). The inclusions even
dominate the [Ne~{\sc ii}] 12.8-$\mu$m line flux.  Note that gas densities in
the H-deficient inclusions are not well constrained at the moment, partly due
to the lack of suitable atomic data and diagnostic tools (see below, \S{1.8}).
This leads to some uncertainties in the deduced total mass of the H-deficient
material. In the current treatment, the inclusions have densities and
temperatures that are roughly in pressure equilibrium with the surrounding
diffuse medium of higher temperatures and lower densities. Physically, one
envisions individual inclusions to be optically thick such that they can
survive long enough to have observational effects. Proper modeling of optically
thick knots will however require a separate set of fine grid for each of them
to resolve the ionization and thermal structures, and thus demand even more
computing resources.

The average elemental abundances for the whole nebula of NGC\,6153, including
the diffuse nebula as well as H-deficient inclusions embedded in it, are 0.102,
$3.9\times 10^{-4}$, $4.4\times 10^{-4}$, $7.3\times 10^{-4}$ and $2.4\times
10^{-4}$ for He, C, N, O and Ne, respectively. Except for helium, elemental
abundances of the ``normal'' component are close to those deduced from the
empirical method based on CELs (Liu et al., 2000), and are about 30 per cent
lower than the average abundances for the whole nebula. For helium, the
abundance of the ``normal'' component is about 40 per cent lower than derived
from the empirical method based on the observed strengths of helium ORLs, and
actually comes closer to the average helium abundance of the whole nebula. That
NGC\,6153 has a lower helium abundance than implied by the empirical analysis
is supported by a measurement of the near IR He~{\sc i} $\lambda$10830 line,
for which excitation from the 2s\,$^3$S meta-stable level by electron impacts
dominates the emission (Yuan et al., 2010).

Helium lines observable in the optical and UV are all dominated by
recombination excitation, and their strengths can be strongly enhanced by a
small amount of H-deficient, ultra-cold plasma posited in the nebula. If
H-deficient inclusions are indeed fully responsible for the ORL versus CEL
dichotomy ubiquitously found in PNe and H~{\sc ii} regions, then their presence
may have a profound consequence on the helium abundances deduced for those
objects, as well as on the primordial helium abundance determined by
extrapolating the helium abundances of metal-poor H~{\sc ii} galaxies to zero
metallicity -- the inferred primordial helium abundances from standard analyses
would be too high (Zhang et al., 2004).

Corroborative evidence that ORLs arise from distinct regions of very low
temperature plasma is provided by analyses of emission line profiles.
Spectroscopy at a resolution of 6\,km\,s$^{-1}$ of NGC\,7009 shows that O~{\sc
ii} ORLs are significantly narrower than the [O~{\sc iii}] forbidden lines
(Liu, 2006b). Observations at an even higher resolution of 2\,km\,s$^{-1}$ of
NGC\,7009 and NGC\,6153 yield complex multi-velocity-components and differing
line profiles for the O~{\sc ii} ORLs and for the [O~{\sc iii}] forbidden
lines.  The profiles of the [O~{\sc iii}] $\lambda$5007 nebular and
$\lambda$4363 auroral lines also differ significantly. These results suggest
that ionized regions of vastly different temperatures (and radial and thermal
velocities) co-exist in those nebulae (Barlow et al., 2006). Zhang (2008) shows
that the differences in the observed line profiles of the O~{\sc ii} ORLs and
the [O~{\sc iii}] CELs are hard to explain in a chemically homogeneous nebula. 

\section{Spectroscopy of PNe harboring H-deficient inclusions}

\begin{figure}
\centering
\centerline{\includegraphics[width=160mm]{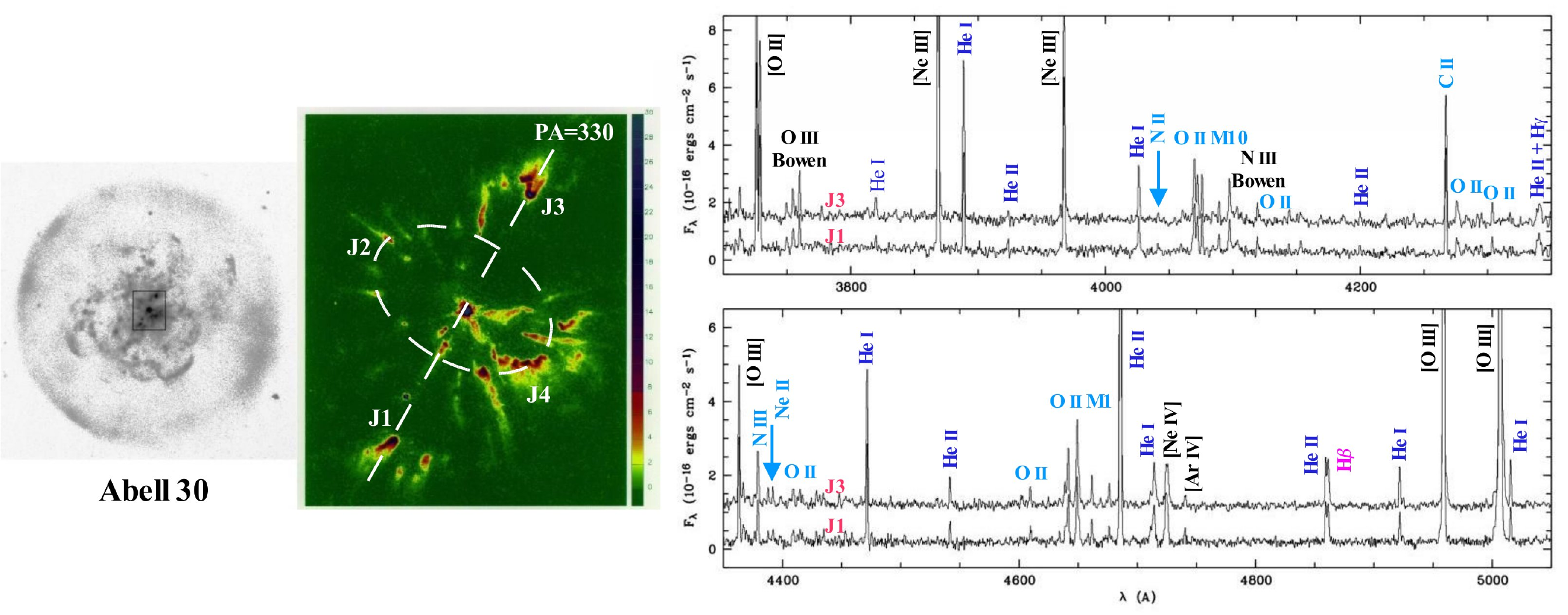}}

\caption{{\em Left}: The optical images of Abell\,30, showing the H-deficient
knots near the nebular center (adapted from Borkowski et al., 1993); {\em
Right}: The optical spectrum of the polar knots J\,1 and J\,3 showing prominent
ORLs of C, N, O and Ne ions (adapted from Wesson et al., 2003). Note the
faintness of H$\beta$ at 4861\,{\AA}.} 

\label{fig07}
\end{figure}

H-deficient clumps were previously known to exist in a rare class of old PNe,
including Abell\,30, 58 and 78, IRAS\,15154-5258 and 18333-2357 (e.g.
Harrington, 1996). They are identified as PNe that experience a last helium
shell flash that brings them back to the AGB to repeat the PN evolution stage,
the so-called ``born-again'' PNe (Iben et al., 1983). In Abell\,30, HST imaging
reveals a host of knots embedded near the center of a round, limb-brightened
faint nebula of angular diameter 127\,arcsec (Borkowski et al., 1993; c.f.
Fig.\,\ref{fig07}). They include two point-symmetric polar knots along ${\rm
PA} = 331^{\rm o}$ at angular distances 6.66 and 7.44\,arcsec from the central
star, plus a number of knots loosely delineating an equatorial disk or ring.
Deep optical spectra of knots J\,1 and J\,3 were obtained by Wesson et al.
(2003).  The spectra show prominent recombination lines from CNONe ions,
remarkably similar to those seen in other PNe of large adf's, such as Hf\,2-2
in Fig.\,\ref{fig04}, except that in Abell\,30 the strengths of those CNONe
ORLs relative to H$\beta$ are much higher. Detailed ORL analyses show that the
ORLs are indeed emitted under very low temperatures of only a few hundred
Kevin. The result is corroborated by detailed 3D photoionization modeling
(Ercolano et al., 2003b) as well as by a similar analysis of the H-deficient
knot found at the center of Abell\,58 by Wesson et al. (2008). 

\section{Origins of H-deficient inclusions}

The origin of H-deficient material is not well understood and their existence
in PNe, and possibly also in H~{\sc ii} regions (see, for example, Esteban et
al. 2002 and Tsamis et al. 2003a), is not predicted by the current theory of
stellar evolution. For PNe harboring H-deficient central stars, such as the
Wolf-Rayet and PG\,1159 stars, while the scenario of a single post-AGB star
experiencing a last helium shell flash seems to be able to match the
photospheric abundances of the star, not all PNe exhibiting large adf's have an
H-deficient central star (e.g. NGC\,7009). The process and mechanism by which
H-deficient knots might be ejected in such systems are not understood.  More
importantly, detailed ORL abundance analyses by Wesson et al.  (2003, 2008)
show that the H-deficient knots found in Abell\,30 and Abell\,58 are
oxygen-rich, not carbon-rich as one expects in the scenario of ``born-again''
PNe.

The precise collinearity of the two polar knots in Abell\,30 with the central
star, to within 5\,arcmin, is hard to explain by single star evolution, and
suggests instead the action of a bipolar jet from, for example, an accretion
disk in a binary system (Harrington, 1996; De Marco, 2008). In this respect, it
is interesting to note that Abell\,58 is known to have experienced a nova-like
outburst in 1917 (Clayton and De Marco, 1997).  Hf\,2-2 has also been found to
be a close binary system with a period of 0.398\,571\,d (Lutz et al., 1998).
By comparing the known properties of Abell\,58 with those of Abell\,30,
Sakurai's Object and several novae and nova remnants, Wesson et al.  (2008)
argue that the elemental abundances in the H-deficient knots of Abell\,30 and
Abell\,58 have more in common with neon novae than with Sakurai's Object, which
is believed to have undergone a final helium flash.


An alternative to the scenario of H-deficient inclusions being ejecta of
nucleo-processed material is that they derive from metal-rich planetary
material, such as icy planetesimals, that once orbited the PN progenitor star.
As the star enters the PN phase by ejecting the envelope and evolves to become
a luminous, hot white dwarf, the strong stellar winds and UV radiation fields
begin to photoionize and strip gas off the planetesimals, now embedded in the
PN (Liu, 2003, 2006a). As analyses show that only a few Jupiter masses
metal-rich material is required to explain the observed strengths of ORLs, the
idea may not be so eccentric as it first looks. Evaporating proto-planetary
disks around newly formed stars (such as the proplyds found in the Orion
Nebula) may likewise provide a natural solution to the problem of ORL versus
CEL dichotomy similarly found in H~{\sc ii} regions.

\section{The need for new atomic data}

Atomic data relevant for the study of emission line nebulae, including
collision strengths and effective recombination coefficients, are generally
calculated for a temperature range 5,000 -- 20,000\,K, typical for photoionized
nebulae of solar composition. The finding that heavy element ORLs emitted by
the H-deficient knots in the ``born-again'' PNe Abell\,30 and Abell\,58 arise
from ultra-cold plasma of only a few hundred Kevin, as well as the compelling
evidence accumulated so far that points to the presence of similar cold
H-deficient inclusions in other PNe, call for new atomic data, in particular
the effective recombination coefficients applicable in such low temperature
environments. New plasma diagnostics that use only ORLs, not CELs, need to be
developed so that physical and chemical properties of the H-deficient
knots/inclusions, such as temperature, density, elemental abundances and mass,
can be determined reliably, a prerequisite to unravel their origins. 

\begin{figure} \centering
\centerline{\includegraphics[width=160mm]{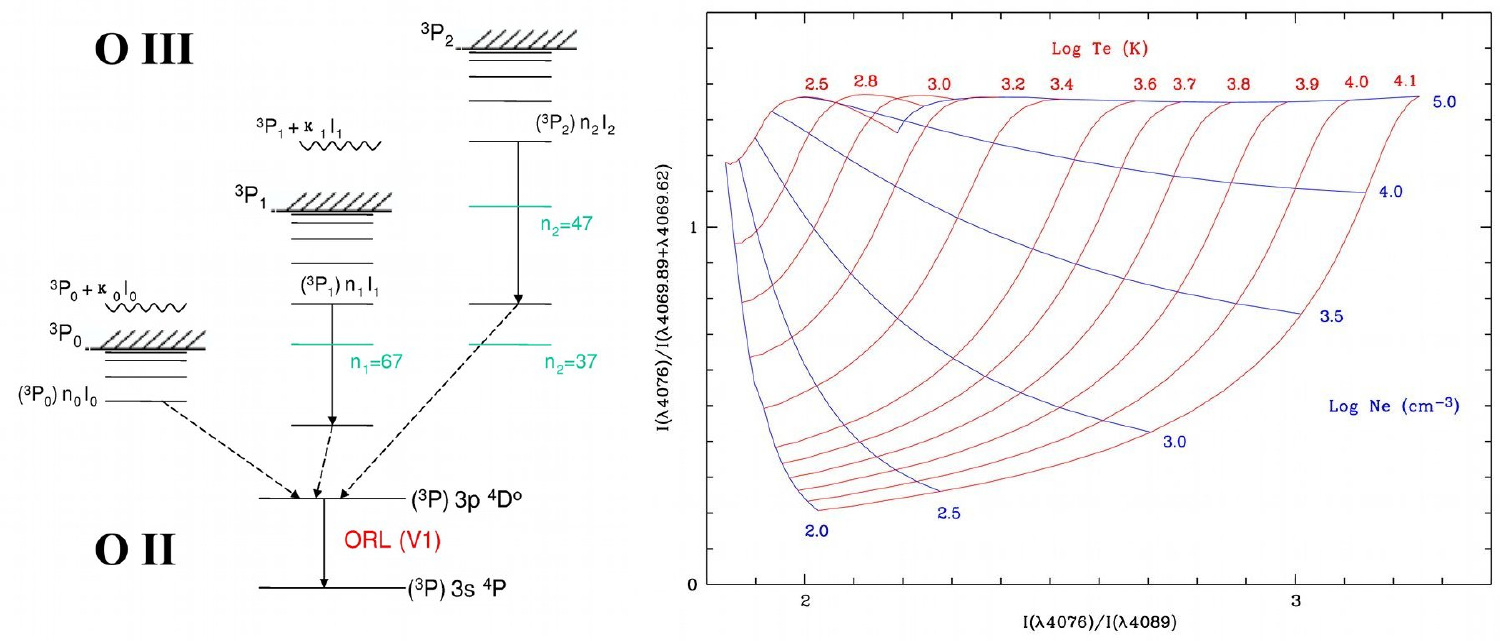}}

\caption{{\em Left}: Schematic diagram illustrating the low temperature
di-electronic recombination of O~{\sc ii} via the low-lying fine-structure
auto-ionizing states.  Note that for $n_1 \ge 67$ and $n_2 \ge 37$, the
doubly-excited O~{\sc ii} 2p$^2$($^3$P$_1$)$n_1 l_1$ and 2p$^2$($^3$P$_2$)$n_2
l_2$ states fall above the O~{\sc iii} ground level 2p$^2$\,$^3$P$_0$.
Similarly, for $n_2 \ge 47$, O~{\sc ii} 2p$^2$($^3$P$_2$)$n_2 l_2$ states have
energies above the O~{\sc iii} first excited fine-structure level
2p$^2$\,$^3$P$_1$. For temperatures of a few hundred Kelvin, di-electronic
recombination via those low-lying fine-structure auto-ionizing states becomes
an important process; {\em Right}: Loci of the O~{\sc ii} recombination line
ratios $\lambda4076/\lambda4089$ and $\lambda4076/\lambda4069$ for different
$T_{\rm e}$'s and $N_{\rm e}$'s (based on data provided by Dr. P. J. Storey).}

\label{fig08}
\end{figure}

The ratios of the intensities of ORLs from states of different orbital angular
momenta show some temperature dependence and thus can be used to measure the
average temperature under which the lines are emitted (Liu, 2003). In addition,
as pointed out by Liu (2003), the dependence on density of the relative
populations of the fine-structure levels of the ground spectral term of the
recombining ions, such as O$^{2+}$ 2p$^2$\,$^3$P$_{0,1,2}$, leads to variations
of the relative intensities of recombination lines as a function of density.
In the case of O~{\sc ii}, for example, lines from high-$l$ states, such as the
3p\,$^4$D$^{\rm o}_{7/2}$ -- 3s\,$^4$P$_{5/2}$ $\lambda$4649 transition (c.f.
Fig.\,\ref{fig02}) weakens as density decreases due to the underpopulation of
the O$^{2+}$ 2p$^2$\,$^3$P$_2$ level under low densities.  This effect opens up
the possibility of determining electron density using ORLs.

New effective recombination coefficients, calculated down to temperatures of
$\sim 100$\,K and taking into account the dependence on density of level
populations of the ground states of the recombining ion, have now been carried
out for the O~{\sc ii} (Bastin and Storey, 2006) and N~{\sc ii} (Fang, Storey
and Liu, in preparation) recombination spectra. Fig.\,\ref{fig08} plots loci of
the O~{\sc ii} $I(\lambda4076)/I(\lambda4089)$ and
$I(\lambda4076)/I(\lambda4070)$ intensity ratios for different temperatures and
densities. The diagram also illustrates the importance of di-electronic
recombination via the low-lying fine-structure auto-ionizing states under very
low temperatures. Preliminary applications of these new atomic data and
diagnostics to observations yield lower temperatures and higher densities than
CELs, as one expects.

\section{Summary}

To summarize, nearly one and half centuries after William Huggins' discovery of
bright emission lines in the spectra of gaseous nebulae, enormous progress has
been achieved in understanding them. In particular, we believe we now
understand the faint ORLs emitted by heavy element ions, more than half a
century after Wyse's pioneer work on O~{\sc ii}. The theory of photoionized
gaseous nebulae seems to be solid. Yet any progress in our understanding of
their nature seems to be always accompanied by the emergence of new puzzles.
Clearly, those beautiful heavenly objects, queens of the night, are more
delicate than we think and still hold some fascinating secrets from us. With
dedicated observations, more discoveries are sure to come.




\newpage
\begin{flushleft}
\textbf{\Large{Document Control Sheet---List of Elements in Chapter}}
\end{flushleft}

\subsubsection{Tables}

\begin{itemize}
\item Table 1:
   \begin{itemize}
   \item \textit{Title:}
   \item \textit{Source/permission status/credit line:}
   \end{itemize}

\item Table 2:
   \begin{itemize}
   \item \textit{Title:}
   \item \textit{Source/permission status/credit line:}
   \end{itemize}

\item Table 3:
   \begin{itemize}
   \item \textit{Title:}
   \item \textit{Source/permission status/credit line:}
   \end{itemize}

\item Table 4:
   \begin{itemize}
   \item \textit{Title:}
   \item \textit{Source/permission status/credit line:}
   \end{itemize}

\item Table 5:
   \begin{itemize}
   \item \textit{Title:}
   \item \textit{Source/permission status/credit line:}
   \end{itemize}

\item Table n, etc.\footnote{Add or delete items as necessary. Include a permission-status statement describing whether copyright permission is needed and, if so, whether it has been obtained. Include a credit line for each table that is reproduced from another source, even from your own previously published work.}
\end{itemize}

\subsubsection{Figures}

\begin{itemize}
\item Figure 1:
   \begin{itemize}
   \item \textit{Label:}
   \item \textit{Filename:}
   \item \textit{Source/permission status/credit line:}
   \end{itemize}

\item Figure 2:
   \begin{itemize}
   \item \textit{Label:}
   \item \textit{Filename:}
   \item \textit{Source/permission status/credit line:}
   \end{itemize}

\item Figure 3:
   \begin{itemize}
   \item \textit{Label:}
   \item \textit{Filename:}
   \item \textit{Source/permission status/credit line:}
   \end{itemize}

\item Figure 4:
   \begin{itemize}
   \item \textit{Label:}
   \item \textit{Filename:}
   \item \textit{Source/permission status/credit line:}
   \end{itemize}

\item Figure 5:
   \begin{itemize}
   \item \textit{Label:}
   \item \textit{Filename:}
   \item \textit{Source/permission status/credit line:}
   \end{itemize}

\item Figure n, etc.\footnote{Add or delete items as necessary. Include a permission-status statement describing whether copyright permission is needed and, if so, whether it has been obtained. Include a credit line for each figure that is reproduced from another source, even from your own previously published work. When submitting your final files for publication, please provide source image files (in .eps or .ps format) for all figures at a minimum resolution of 300 dpi.}
\end{itemize}


\end{doublespace}

\newpage

\begin{thebibliography}{99}

\bibitem{am1945}
Aller, L. H. and Menzel, D. H. (1945). Physical Processes in Gaseous Nebulae. XVIII. The chemical composition of the planetary 
nebulae. ApJ, 102, 239

\bibitem{abm1939}
Aller, L. H., Baker, J. G. and Menzel D. H. (1939). Physical processes in 
gaseous nebulae. VIII. The ultraviolet radiation field and electron 
temperature of an optically thick nebula. ApJ, 90, 601

\bibitem{bm1938}
Baker, J. G. and Menzel, D. H. (1938). Physical processes in gaseous nebulae. 
III. The Balmer decrement. ApJ, 88, 52;

\bibitem{bma1938}
Baker, J. G., Menzel, D. H. and Aller, L. H. (1938). Physical processes in 
gaseous nebulae. V. Electron temperatures. ApJ, 88, 422

\bibitem{barker1978}
Barker, T. (1978). Spectrophotometry of planetary nebulae. I - Physical 
conditions. ApJ, 219, 914

\bibitem{barker1991}
Barker, T. (1991). The ionization structure of planetary nebulae. X -- 
NGC\,2392. ApJ, 371, 217

\bibitem{Barlow03}
Barlow, M. J., Liu, X.-W., P\'{e}quignot, D., Storey, P. J., Tsamis, Y., 
Morisset, C.,  (2003). PN recombination line abundances for magnesium, 
silicon and sulphur. In Planetary Nebulae: Their Evolution and Role in the 
Universe, Proc.  IAU Symp. \#209. Eds. S. Kwok, M. Dopita and R. Sutherland. 
PASP. pp.373-374

\bibitem{barlow06}
Barlow, M. J., Hales, A. S., Storey, P. J., Liu, X.-W., Tsamis, Y. G. and
Aderin, M. E. (2006). bHROS high spectral resolution observations of PN 
forbidden and recombination line profiles planetary nebulae. In Planetary 
Nebulae in our Galaxy and Beyond, Proceedings of the IAU Symposium \#234. 
Eds. M. J. Barlow and R. H. M\'{e}ndez Cambridge: Cambridge University Press. 
p.367

\bibitem{bastin06}
Bastin, R. J. and Storey, P. J. (2006). Recombination line spectroscopy: 
the O~{\sc ii} spectrum. In Planetary Nebulae in our Galaxy and Beyond, 
Proceedings of the IAU Symposium \#234. Eds. M. J. Barlow and 
R. H. M\'{e}ndez. Cambridge: Cambridge University Press. p.369

\bibitem{borkowski93}
Borkowski, K. J., Harrington, J. P., Tsvetanov, Z., Clegg, R. E. S. (1993).
HST imaging of hydrogen-poor ejecta in Abell 30 and Abell 78 - Wind-blown 
cometary structures. ApJ, 415, 47

\bibitem{bowen1927a}
Bowen, I. S. (1927a). Series spectra of boron, carbon, nitrogen, oxygen and 
fluorine. Phys.Rev., 29, 231.

\bibitem{bowen1927b}
Bowen, I. S. (1927b). The Origin of the Nebulium Spectrum. Nature, 120, 473

\bibitem{bowen1927c}
Bowen, I. S. (1927c). The origin of the chief nebular lines. PASP, 39, 295


\bibitem{bowen1934}
Bowen, I. S. (1934). The Excitation of the Permitted O III Nebular Lines. 
PASP, 46, 146

\bibitem{bowen1935}
Bowen, I. S. (1935). The Spectrum and Composition of the Gaseous Nebulae.
ApJ, 81, 1

\bibitem{bowen1938}
Bowen, I. S. (1938). The image-slicer, a device for reducing loss of light 
at slit of stellar spectrograph. ApJ, 88, 113

\bibitem{bw39}
Bowen, I. S. and Wyse, A. (1939). The spectra and chemical composition of the
gaseous nebulae, NGC 6572, 7027, 7662. Lick Obs. Bull., 19, 1 

\bibitem{burgess1958}
Burgess, A. (1958). The hydrogen recombination spectrum. MNRAS, 118, 477

\bibitem{clayton97}
Clayton, G. C. and De Marco, O. (1997). The evolution of the final helium shell flash 
star V\,605 Aquilae from 1917 to 1997. AJ, 114, 2679

\bibitem{cunto93}
Cunto, W., Mendoza, C., Ochsenbein, F. and Zeippen, C. J. (1993). Topbase at
the CDS. A\&A, 275, 5

\bibitem{DeMarco08}
De Marco, O. (2008). [WC] and PG 1159 central stars of planetary nebulae: the 
need for an alternative to the born-again scenario. In Hydrogen-Deficient Stars.
Eds. K. Werner and T. Rauch. PASP: San Francisco. p.209


\bibitem{ercolano03a}
Ercolano, B., Barlow, M. J., Storey, P. J. and Liu, X.-W. (2003a). MOCASSIN: 
a fully three-dimensional Monte Carlo photoionization code. MNRAS, 340, 1136

\bibitem{ercolano03b}
Ercolano, B., Barlow, M. J., Storey, P. J., Liu, X.-W., Rauch, T. and
Werner K. (2003b). Three-dimensional photoionization modelling of the 
hydrogen-deficient knots in the planetary nebula Abell\,30. MNRAS, 344, 1145

\bibitem{esteban02}
Esteban, C., Peimbert, M., Torres-Peimbert, S. and Rodr\'{i}guez, M. (2002).
Optical recombination lines of heavy elements in giant extragalactic 
H\,{\sc ii} regions. ApJ, 581, 241

\bibitem{flower1968}
Flower, D. R. (1969). The ionization structure of planetary nebulae -- VII. The 
heavy elements. MNRAS, 146, 171; The ionization structure of planetary 
nebulae -- VIII. Models of the nebulae NGC 7662 and IC 418. ibid, 146, 243

\bibitem{goodson1967}
Goodson, W. L. (1967). Ionisationsstruktur planetarischer nebel. Z.Astrophys., 
66, 118

\bibitem{harringtn1968}
Harrington, J. P. (1968). Ionization stratification and thermal stability in 
model planetary nebulae. ApJ, 152, 943

\bibitem{Harrington96}
Harrington, J. P. (1996). Observations and models of H-deficient planetary 
nebulae. In Hydrogen-Deficient Starsi, ASP Conf.  Ser. Vol. 96. eds. C. S. 
Jeffery, U. Hebera. PASP: San Francisco. p.193

\bibitem{Herschel1785}
Herschel, W. (1785), On the construction of the heavens. 
Phil.Trans.R.Soc.London, 75, 213.

\bibitem{hjellming1966}
Hjellming, R. (1966). Physical processes in H II regions. ApJ, 143, 120


\bibitem{huggins1882}
Huggins, W. (1882). Note on the photographic spectrum of the Great Nebula in 
Orion. Proc. R. Soc. London, 33, 425

\bibitem{hm1964}
Huggins, W. and Millar, W. A. (1864). On the spectra of some of the fixed 
stars. Phil. Trans. R. Soc. London, 154, 437

\bibitem{iben83}
Iben, I., Jr., Kaler, J. B., Truran, J. W. and Renzini, A. (1983). On the
evolution of those nuclei of planetary nebulae that experience a final 
helium shell flash. ApJ, 264, 605


\bibitem{kaler1966}
Kaler, J. B. (1966). Hydrogen and Helium Spectra of Gaseous Nebulae. ApJ, 
143, 722

\bibitem{kb1960}
Kirchhoff, G. and Bunsen, R. W. (1860). Chmical analysis by obseration of 
spectra. Annalen der Physik und der Chemie (Poggendorff), Vol. 110, pp.161-189.

\bibitem{liu03} 
Liu, X.-W. (2003). Probing the Dark Secrets of PNe with ORLs (invited review).
In Planetary Nebulae: Their Evolution and Role in the Universe, Proc.  IAU
Symp. \#209. Eds. S. Kwok, M. Dopita and R. Sutherland. PASP. pp.339-346

\bibitem{liu06a}
Liu, X.-W. (2006a). Optical recombination lines as probes of conditions in 
planetary nebulae. In Planetary Nebulae in our Galaxy and Beyond, Proceedings 
of the IAU Symposium \#234. Eds. M. J. Barlow and R. H. M\'{e}ndez.
Cambridge: Cambridge University Press. pp.219-226

\bibitem{liu06b}
Liu, X.-W. (2006b). Plasma diagnostics and elemental abundance determinations 
for PNe current status. In Planetary nebulae beyond the Milky Way, eds. J. 
Walsh, L. Stanghellini and N. Douglas, p.169

\bibitem{lbzb06}
Liu, X.-W., Barlow, M. J., Zhang, Y., Bastin, R. J. and Storey, P. J. (2006).
Chemical abundances for Hf 2-2, a planetary nebula with the strongest-known 
heavy-element recombination lines. MNRAS, 368, 1959

\bibitem{lbcd01}
Liu, X.-W., Barlow, M. J., Cohen, M., Danziger, I. J., Luo, S.-G., 
Baluteau, J. P., Cox, P., Emery, R. J., Lim, T. and P\'{e}quignot, D. (2001a).
ISO LWS observations of planetary nebula fine-structure lines. MNRAS, 323, 343

\bibitem{ld93}
Liu, X.-W. and Danziger, I. J. (1993). Electron temperature determination from 
nebular continuum emission in planetary nebulae and the importance of 
temperature fluctuations. MNRAS, 263, 256

\bibitem{lldb01}
Liu, X.-W., Luo, S.-G., Barlow, M. J., Danziger, I. J. and Storey, P. J. 
(2001b). Chemical abundances of planetary nebulae from optical recombination 
lines -- III. The Galactic bulge PN M\,1-42 and M\,2-36

\bibitem{lsbc95}
Liu, X.-W., Storey, P. J., Barlow, M. J. and Clegg, R. E. S. (1995).
The rich O II recombination spectrum of the planetary nebula NGC 7009: new 
observations and atomic data. MNRAS, 272, 369

\bibitem{lsbd00}
Liu, X.-W., Storey, P. J., Barlow, M. J., Danziger, I. J., Cohen, M. and 
Bryce, M. (2000). NGC 6153: a super-metal-rich planetary nebula? MNRAS, 312, 
585

\bibitem{liuy04a}
Liu, Y., Liu, X.-W., Luo, S.-G. and Barlow, M. J. (2004a). 
Chemical abundances of planetary nebulae from optical recombination lines -- 
I. Observations and plasma diagnostics. MNRAS, 353, 1231

\bibitem{liuy04b}
Liu, Y., Liu, X.-W., Barlow, M. J. and Luo, S.-G. (2004b).
Chemical abundances of planetary nebulae from optical recombination lines -- 
II. Abundances derived from collisionally excited lines and optical 
recombination lines. MNRAS, 353, 1251

\bibitem{luo01}
Luo, S.-G., Liu, X.-W. and Barlow, M. J. (2001). Chemical abundances of 
planetary nebulae from optical recombination lines -- II. The neon 
abundance of NGC\,7009. MNRAS, 326, 1049

\bibitem{lutz98}
Lutz, J., Alves, D. and Becker, A., et al. (1998). 
V and R magnitudes for planetary nebula central stars in the MACHO project 
galactic bulge fields. AAS, 192, 5309

\bibitem{ml99}
Mathis, J. S. and Liu, X-W. (1999). Observations of the [O~{\sc iii}] 
$\lambda$4931/$\lambda$lambda4959 line ratio and O$^{+2}$ abundances in 
ionized nebulae. ApJ, 521, 212

\bibitem{ma1941}
Menzel, D. H. and Aller, L. H. (1941). Physical processes in gaseous nebulae. 
XVI. The abundance of O~{\sc iii}. ApJ, 94, 30

\bibitem{mah1941}
Menzel, D. H., Aller, L. H. and Hebb, M. H. (1941). Physical processes in 
gaseous nebulae. XIII. ApJ, 93, 230.

\bibitem{miller1971}
Miller, J. S. (1971). Photoelectric measurements of high-n Balmer lines in 
NGC\,7027 and NGC\,7662. ApJ, 165, L101

\bibitem{miller1971}
Miller, J. S. and Mathews, W. G. (1972). The recombination spectrum of the 
planetary nebula NGC\,7027. ApJ, 172, 593

\bibitem{osterbrock1974}
Osterbrock, D. E. (1974). Astrophysics of Gaseous Nebulae.  Sausalito: 
University Science Books.

\bibitem{of05}
Osterbrock, D. E. and Ferland, G. J. (2005). Astrophysics of Gaseous Nebulae
and Active Galactic Nuclei. Sausalito: University Science Books.

\bibitem{pagel97}
Pagel, B. E. J. (1997). Nucleosynthesis and Chemical Evolution of Galaxies.
Cambridge: Cambridge University Press

\bibitem{peimbert1967}
Peimbert, M. (1967). Temperature determinations of H~{\sc ii} regions. ApJ, 
150, 825

\bibitem{peimbert1971}
Peimbert, M. (1971). Planetary nebulae II. Electron temperatures and electron
densities. Bol.Obs.Ton.Tac., 6, 29

\bibitem{pequignot02}
P\'{e}quignot, D., Amara, M., Liu, X.-W., Barlow, M. J., Storey, P. J., 
Morisset, C., Torres-Peimbert, S. and Peimbert, M. (2002). Photoionization 
models for planetary nebulae with inhomogeneous chemical composition.
RMxAC, 12, 142

\bibitem{rola1994}
Rola, C. and Stasi\'{n}ska, G. (1994). The carbon abundance problem in 
planetary nebulae. A\&A, 282, 199

\bibitem{rubin1968}
Rubin, R. H. (1968). The Structure and Properties of H II Regions. ApJ, 153, 761

\bibitem{russell1927}
Russell, H. N., Dugan, R. S., Stewart, J. Q. (1927). Astronomy. 
Boston: Ginn and Co.

\bibitem{schmidt1932} 
Schmidt B. (1932). Mitteilungen Hamburger Sternwarte in Bergedorf 7(36), 15.
Reprinted: Selected Papers on Astronomical Optics, SPIE Milestone Ser. 73, 165
(1993)

\bibitem{seaton1954}
Seaton, M. J. (1954). Electron temperatures and electron densities in 
planetary nebulae. MNRAS, 114, 154

\bibitem{seaton1959}
Seaton, M. J. (1959). Radiative recombination of hydrogenic ions. MNRAS, 119, 
81; The solution of capture-cascade equations for hydrogen. ibid, 119, 90 

\bibitem{so1959}
Seaton, M.  J. and Osterbrock, D. E. (1957). Relative [O~{\sc ii}] intensities 
in gaseous nebulae. ApJ, 125, 66
 
\bibitem{seaton1960}
Seaton, M. J. (1960). H~{\sc i}, He~{\sc i}, and He{\sc ii} intensities in 
planetary nebulae. MNRAS, 120, 326

\bibitem{seaton1987}
Seaton, M. J. (1987). Atomic data for opacity calculations: I. General 
description. J.Phys.B, 20, 6363

\bibitem{shklovsky1956}
Shklovsky I. S. (1956). Astron. Zh., 33, 315

\bibitem{spitzer1948}
Spitzer, L. (1948). The temperature of interstellar matter. I. ApJ, 107, 6

\bibitem{stromgren1939}
Str\"{o}mgren, B. (1939). The physical state of interstellar hydrogen. 
ApJ, 89, 526

\bibitem{tsamis03a}
Tsamis, Y. G., Barlow, M. J., Liu, X.-W., Danziger, I. J. and Storey, P. J.
(2003a). Heavy elements in Galactic and Magellanic Cloud H~{\sc ii} regions: 
recombination-line versus forbidden-line abundances. MNRAS, 338, 687

\bibitem{tsamis03b}
Tsamis, Y. G., Barlow, M. J., Liu, X.-W., Danziger, I. J. and Storey, P. J.
(2003b). A deep survey of heavy element lines in planetary nebulae -- I. 
Observations and forbidden-line densities, temperatures and abundances.
MNRAS, 345, 186

\bibitem{tsamis04}
Tsamis, Y. G., Barlow, M. J., Liu, X.-W., Storey, P. J. and Danziger, I. J.
(2004). A deep survey of heavy element lines in planetary nebulae -- II. 
Recombination-line abundances and evidence for cold plasma.
MNRAS, 353, 953

\bibitem{wang07}
Wang, W. and Liu, X.-W. (2007). Elemental abundances of Galactic bulge 
planetary nebulae from optical recombination lines. MNRAS, 381, 669

\bibitem{wesson08}
Wesson, R., Barlow, M. J., Liu, X.-W., Storey, P. J., Ercolano, B. and 
de Marco, O. (2008). The hydrogen-deficient knot of the `born-again' planetary 
nebula Abell\,58 (V605 Aql). MNRAS, 383, 1639

\bibitem{wesson03}
Wesson, R., Liu, X.-W. and Barlow, M. J. (2003). Physical conditions in the 
planetary nebula Abell\,30. MNRAS, 340, 253

\bibitem{wesson05}
Wesson, R., Liu, X.-W. and Barlow, M. J. (2005). The abundance discrepancy -- 
recombination line versus forbidden line abundances for a northern sample of 
galactic planetary nebulae
MNRAS, 362, 424

\bibitem{wood1910}
Wood R. (1910). The echellette grating for the infra-red. Philos. Mag. 20 
(series 6), 770-778

\bibitem{wright1918}
Wright W. H. (1918). The wave lengths of the nebular lines and general
observations of the spectra of the gaseous nebulae. Publ. Lick Obs., 13, 191

\bibitem{wyse1942}
Wyse, A. B. (1942). The Spectra of Ten Gaseous Nebulae. ApJ, 95, 356

\bibitem{zanstra1926}
Zanstra, H. (1926). An application of the quantum theory to the luminosity of 
diffuse nebulae. Phys.Rev., 27, 644

\bibitem{zanstra1927}
Zanstra, H. (1927). An application of the quantum theory to the luminosity of 
diffuse nebulae. ApJ, 65, 50

\bibitem{zanstra1931}
Zanstra, H. (1931). Luminosity of planetary nebulae and stellar temperatures.
Publ.Dom.Astrophys.Obs. (Victoria), 4, 209

\bibitem{zhang04}
Zhang, Y., Liu, X.-W., Wesson, R., Storey, P. J., Liu, Y. and Danziger, I. J.
(2004). Electron temperatures and densities of planetary nebulae determined 
from the nebular hydrogen recombination spectrum and temperature and density 
variations. MNRAS, 351, 935

\bibitem{zhang05}
Zhang, Y., Liu, X.-W., Liu, Y. and Rubin, R. H. (2005). Helium recombination 
spectrum as temperature diagnostics for planetary nebulae MNRAS, 358, 457

\bibitem{zhang08}
Zhang, Y. (2008). Emission line profiles as a probe of physical conditions in 
planetary nebulae. A\&A, 486, 221

\end{thebibliography}
\end{document}